 \documentclass[prl,aps,twocolumn,showpacs]{revtex4}
\usepackage[dvips]{graphicx}
\usepackage{dcolumn}%
\usepackage{amsmath}
\usepackage{amsfonts}%
\usepackage{amssymb}
\providecommand{\U}[1]{\protect\rule{.1in}{.1in}}
\newcommand{\be}{\begin{equation}}
\newcommand{\ee}{\end{equation}}
\newcommand{\bea}{\begin{eqnarray}}
\newcommand{\eea}{\end{eqnarray}}
\newcommand{\bt} {\begin{tabular}}
\newcommand{\et} {\end{tabular}}
\newcommand{\nn}{ \nonumber}

\newcommand{\ba} {\begin{array}}
\newcommand{\ea} {\end{array}}
\begin{document}

\title{Scattering theory of   thermocurrent in quantum dots and molecules}

\author{Natalya A. Zimbovskaya}

\affiliation
{Department of Physics and Electronics, University of Puerto
Rico-Humacao, CUH Station, Humacao, Puerto Rico 00791, USA,  \\
Institute for Functional Nanomaterials, University of Puerto \\ Rico,
San Juan, Puerto Ruco 00931, USA \\
{\bf Email}: natalya.zimbovskaya@upr.edu}

\begin{abstract}
In this work we theoretically study properties of electric current
driven by a temperature gradient through a quantum dot/molecule
coupled to the source and drain charge reservoirs. We analyze the
effect of Coulomb interactions between electrons on the dot/molecule
and of thermal phonons associated with the electrodes  thermal
environment on the thermocurrent.
 The scattering matrix formalism is employed to compute electron
transmission through the system. This approach is further developed
and combined with nonequilibrium Green's functions formalism, so that
scattering probabilities are expressed in terms of relevant energies
including the  thermal energy, strengths of coupling between the
dot/molecule and charge reservoirs and characteristic energies of
electron-phonon interactions in the electrodes. It is shown that one
may bring the considered system into regime favorable for
heat-to-electric energy conversion by varying the applied bias and
gate voltages.
\vspace{2mm}

   \end{abstract}


\date{\today}
\maketitle

\section{i. introduction}

The field of molecular electronics has been rapidly expanding during
the last two decades owing to continuous improvement of techniques
intended to electrically contact and control quantum dots and single
molecules in transport junctions. Recent advances  in heat
measurements in nanoscale systems allow to  study thermoelectric
properties of molecular junctions and similar systems. These studies
bring a deeper understanding of transport mechanisms \cite{1,2,3,4}
and additional information concerning electronic and vibrational
excitation spectra of molecules \cite{5,6}. Also, in the recent years
a new field of molecular thermoelectronics have emerged \cite{7}.
Thermal analogs of molecular transistors and heat-into-electricity
converters were proposed \cite{5,8,9,10,11,12}.

Heat-to-electric power converters operate due to Seebeck effect which
appears provided that thermal and electric driving forces
simultaneously affect electron transport through the considered
system. When a temperature gradient $ \Delta T $ is applied across the
system, a thermovoltage $ V_{th} $  emerges under the condition of
zero net current thus indicating the energy conversion. At small
temperature gradients $(\Delta T \ll T_0,\ T_0$ being the temperature
of the cool region)  the system operates within the linear regime and
$ V_{th} = - S\Delta T. $ Within this regime, thermopower $ S $ is the
decisive quantity determining the extent of energy conversion.
Accordingly, the thermopower was intensively studied
\cite{5,10,13,14,15,16} along with the thermoelectric figure of merit
\cite{15,17,18}. Another interesting quantity characterizing
thermoelectric transport through molecular junctions and other systems
of similar kind is thermocurrent $ I_{th} . $  The latter may be
defined as a difference between the electron tunnel current flowing
through a biased thermoelectric  junction where the electrodes are
kept at different temperatures  $ (T_L = T_0 + \Delta T,\ T_R = T_0)$
and the current flowing in absence of temperature gradient \cite{19}:
\be
 I_{th} = I(V,T_0,  \Delta T) - I(V,T_0, \Delta T = 0).   \label{1}
\ee
As well as the thermovoltage, the thermocurrent is simultaneously
controlled by  electric and thermal driving forces. The combined
effect of these forces depends of the type of charge carriers involved
in transport and of the bias voltage polarity. When the bias voltage
is sufficiently strong, its effect predominates. Consequently, the
difference between two terms in the Eq. (\ref{1}) diminishes, and $
I_{th} $ approaches zero. At the same time, the thermocurrent flowing
through an unbiased or slightly biased molecule (or quantum dot) is
mostly  controlled by the applied temperature gradient $ \Delta T. $
Assuming that $ V = 0 $ and $ \Delta T > 0 $ (the left electrode is
warmer than the right one), $ I_{th} $ takes on positive/negative
values when the charge carriers involved in transport process are,
respectively, holes/electrons. Finally, when $ V $ and $ \Delta T $
influence the transport to a similar extent, the resulting magnitude
and direction of $ I_{th} $ are determined by both $ \Delta T $ sign
and the bias voltage polarity. These two factors may cooperate or
counteract by pushing charge carriers in the same or opposite
directions. In the latter case, $ I_{th} $may change its sign at
certain values of $V$ and $ \Delta T $ when electric and thermal
driving forces counterbalance each other.
 It was shown that (disregarding electron-phonon
interactions) maximum efficiency of molecular heat-to-electricity
converter could be reached under the condition of vanishing current  $
 I(V,T_0,\Delta T).$  Under this condition, $ I_{th} = -
I(V,T_0,\Delta T = 0).$ Therefore, relatively large in magnitude
thermocurrent flowing through a system may indicate that the system
operates in the regime  favorable for energy conversion \cite{20}.

Properties of the thermovoltage in molecular junctions and quantum
dots were theoretically analyzed in numerous works
\cite{20,21,22,23,24,25,26,27,28,29}.
  As yet, less attention was paid to studies of thermocurrent in spite
of the fact that $I_{th} $ is more straightforward to measure and
model than $ V_{th} ,$ as stated in the recent work \cite{19}.
However, theoretical analysis of $ I_{th} $ behavior within a weakly
nonlinear regime was recently suggested \cite{25}.  The purpose of the
present work is to analyze the effect of the gate voltage,
electron-electron interactions and thermal phonons on the electrodes
on the characteristics of thermocurrent flowing through a
single-molecule junction or a quantum dot.

To properly  analyze thermoelectric transport through
molecules/quantum dots one needs to use an approach including unified
treatment of electron and phonon dynamics in the considered system.
For this purpose, one may use diagrammatic technique or nonequilibrium
Green's function formalism (NEGF), as described in the review
\cite{30}. However, application of these advanced formalisms to
realistic models simulating molecular junctions is extremely
difficult. Several simplified approaches based  on scattering theory
\cite{6} and on quantum rate equations  \cite{5,14,20,31,32} were
developed and used to study thermoelectric properties of molecular
junctions taking into account contributions of vibrational phonons and
electron-vibron interactions. Very recently, a scattering theory based
approach was suggested to analyze weakly nonlinear thermoelectric
transport in mesoscopic systems \cite{25,26,27}. Nevertheless, these
studies are not completed so far.

  In the present work we use a scattering theory first suggested by
Buttiker \cite{33} combined with certain NEGF based results. The
adopted approach allows to derive the expression for the electron
transmission which remains applicable for an arbitrary value of the
difference between the temperatures of the electrodes. Therefore, this
expression may be employed to analyze nonlinear effects in
thermoelectric properties of considered systems.

\section{ii. model and results}

The schematics of the suggested model is presented in the Fig. 1.  To
simplify the computations we mimic the  bridge linking the electrodes
by a single state with the energy $ E_0 .$
 This energy is independent of the electrodes chemical potentials.
However, if the third (gate) electrode is attaches to the system, one
may shift the position of the  bridge level by applying the gate
voltage. We assume that electrons tunnel from the electrodes to the
bridge and vice versa via the channels 1 and 2 thus maintaining an
elastic component of the charge flow. Incoming wave amplitudes $
a_1,a_2 $ and outgoing wave amplitudes $a_1', a_2' $ characterize this
process. Also, electrons in the electrodes may interact with thermal
phonons. An electron may enter the bridge by emitting or absorbing a
phonon. Such electrons contribute to inelastic component of the charge
flow. We assume that the total current is the sum of the elastic and
inelastic components. To describe the inelastic contribution to the
current we introduce a pair of dephasing/dissipative electron
reservoirs.
The channels 3,4 and/or 5,6 connect these reservoirs with the bridge.
While in a reservoir, the electron undergoes scattering by thermal
phonons and then it tunnels to the bridge, so there is no direct coupling of
the bridge to the thermal phonons.  In the present analysis we assume
that the thermal phonons are associated with the left and right
electrodes which are kept at the temperatures $ T_L $ and $T_R. $ The labels indicated the reservoirs are accordingly chosen.

\begin{figure}[t] 
\begin{center}
\includegraphics[width=4.6cm,height=6.9cm,angle=-90]{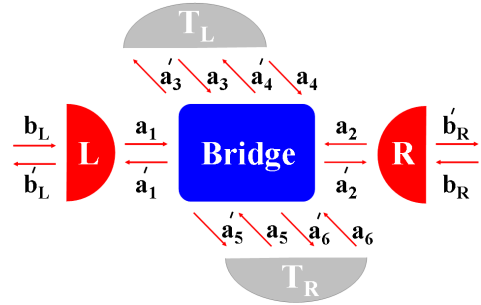}
\caption{(Color online) Schematics of the considered system.
Semicircles  represent the left and right electrodes, square stands
for the molecule/quantum dot sandwiched in between.
Dephasing/dissipative reservoirs are associated with the electrodes
and characterized by temperatures $ T_L $ and $ T_R, $ respectively.
}
 \label{rateI}
\end{center}\end{figure}

So, within the accepted model we have six transport channels. The
relations between incoming particle fluxes $ J_k $ and outgoing fluxes
$ J_k' $ take on the form $(1 \leq i \leq k \leq 6):$
\be
J_k' = \sum T_{ik} J_k.  \label{2}
\ee
Here, the coefficients $ T_{ik} $ are related to matrix elements of
the scattering matrix $ M:\ T_{ik} = |M_{ik}|^2. $ To maintain the
charge conservation in the system,  zero net current should flow in
the channels linking the bridge with the reservoirs:
\begin{align}
J_3 + J_4 - J_3' - J_4' = 0,
\nn\\
J_5 + J_6 - J_5' - J_6' = 0.  \label{3}
\end{align}

The scattering matrix express outgoing wave amplitudes $ b_L', b_R',
a_3',a_4', a_5' $ and $a_6' $ in terms of incident ones $ b_L, b_R,
a_3,a_4, a_5, a_6 .$ To derive expressions for the matrix elements $
M_{ik} $ we first consider a part of the whole system shown in the
Fig. 1. This part consists of the reservoir of the Temperature $ T_L $
associated with the left electrode and the bridge site. Following the
Buttuker's approach \cite{33}, we find the expression far the matrix
$s_{(1)} $ relating wave amplitudes $ a_1', a_2', a_3', a_4'$ to wave
amplitudes  $ a_1, a_2, a_3, a_4 $ namely
\be
s^{(1)} = \left( \ba{cccc}
0  & \sqrt{1 - \epsilon_L} & \sqrt {\epsilon_L} &  0
\\
 \sqrt{1 - \epsilon_L} & 0  & 0 & \sqrt {\epsilon_L}
\\
 \sqrt{ \epsilon_L} & 0  & 0 & - \sqrt {1-\epsilon_L}
\\
0  & \sqrt{\epsilon_L} & -\sqrt {1-\epsilon_L} &  0
\\
\ea  \right) \label{4}
\ee
 Here, $ \epsilon_L $ is the scattering probability associated with
the effect of thermal phonons concentrated on the left electrode (and
represented by the corresponding dephasing/dissipative reservoir).
When $ \epsilon_L = 0, $ the reservoir is detached from the bridge, so
thermal phonons do not affect electron's tunneling from the left
electrode to the bridge. Within the opposite limit $(\epsilon_L =1),$
electrons traveling from the left electrode to the bridge are
necessarily scattered into the reservoir which results in the overall
phase randomization and inelastic transport. Electron tunneling
through a barrier separating the left electrode from the bridge is
characterized by the transmission and reflection amplitudes $(t_L $
and $r_L ,$ respectively). These are matrix elements of a $ 2 \times 2
$ matrix:
 \be
s_L = \left( \ba{cc}
t_L & - r_L \\  r_L & t_L \ea \right)  . \label{5}
\ee
which relates $ b_L',a_1' $ to $ b_L, a_1. $ Combining Egs. (\ref{4})
and (\ref{5}) one obtained the following expression for the matrix $
M^{(1)} $ which relates $ b_L', a_2', a_3',a_4' $ to $ b_L, a_2, a_3,
a_4 :$
\be
M^{(1)} = \left(\ba{cccc}
r_L & \alpha_Lt_L & \beta_Lt_L & 0
\\
\alpha_Lt_L & \alpha_L^2r_L & \alpha_L\beta_Lr_L & \beta_L
\\\beta_Lt_L & \alpha_L\beta_Lr_L & \beta_L^2r_L & -\alpha_L
\\
0 & \beta_L & -\alpha_L & 0  \ea \right).  \label{6}
\ee
Here, $ \alpha_L = \sqrt{1 - \epsilon_L},\ \beta_L =
\sqrt{\epsilon_L}. $ now, we take into consideration the remaining
part of the considered system. As shown in an earlier work \cite{34},
the matrix $M^{(2)} $ relating $a_2', b_R', a_5', a_6' $ to  $a_2,
b_R, a_5, a_6 $ has the form:
\be
M^{(2)} = \left( \ba{cccc}
\alpha_R^2 r_R & \alpha_R t_R & \beta_R & \alpha_R \beta_R r_R
\\
\alpha_R t_R & r_R & 0 & \beta_R t_R
\\
\beta_R & 0 & 0 & -\alpha_R
\\
\alpha_R \beta_R r_R & \beta_R & - \alpha_R & \beta_R^2 r_R
\ea \right) \label{7}
\ee
where $ \alpha_R = \sqrt{1 - \epsilon_R},\ \beta_R =
\sqrt{\epsilon_R}, $ the scattering probability $ \epsilon_R $ is
associated with the effect of thermal phonons concentrated on the
right electrode, and the transmission $(t_R) $ and reflection $(r_R) $
amplitudes characterize electron tunneling between the bridge site and
tunneling between the bridge site the right electrode. Using Eqs.
(\ref{6}), (\ref{7}), one arrives at the following expression for the
scattering matrix \cite{35}:
\begin{widetext}
\be
M = \frac{1}{Z}  \left \{\ba{cccccc}
r_L + \alpha_L^2\alpha_R^2r_R  &\alpha_L \alpha_R t_L t_R   &
\beta_L t_L    &  \alpha_L\alpha_R^2 \beta_L t_L r_R   &
\alpha_L  \beta_R t_L   &  \alpha_L \alpha_R \beta_R t_L r_R
\\
\alpha_L \alpha_R t_L t_R   &  r_R + \alpha_L^2 \alpha_R^2 r_L   &
\alpha_L \alpha_R \beta_L t_R r_L   &  \alpha_R \beta_L t_R &
\alpha_L^2 \alpha_R \beta_R t_R r_L  &  \beta_R t_R
\\
\beta_L t_L  & \alpha_L \alpha_R \beta_L t_R r_L  &
\beta_L^2 r_R  & \alpha_L(\alpha_R^2 r_L r_R -1) &
\alpha_L \beta_L \beta_R r_L  &  \alpha_L \alpha_R \beta_L \beta_R r_L  r_R
 \\
\alpha_L\alpha_R^2 \beta_L t_L r_R  & \alpha_R \beta_L t_R &
\alpha_L(\alpha_R^2 r_L r_{R}-1) & \alpha_R^2 \beta_L^2 r_R &
\beta_L \beta_R   &  \alpha_R \beta_L \beta_R r_R
 \\
\alpha_L \beta_R t_L  & \alpha_L^2 \alpha_R \beta_R t_R r_L &
\alpha_L \beta_L \beta_R r_L  & \beta_L \beta_R &
\alpha_L^2 \beta_R^2 r_L  & \alpha_R (\alpha_L^2 r_L r_R -1)
\\
\alpha_L \alpha_R \beta_R r_R t_L & \beta_R t_R &
\alpha_L\alpha_R \beta_L \beta_R r_L r_R & \alpha_R \beta_L \beta_R
r_R & \alpha_R(\alpha_L^2 r_L r_R-1) & \beta_R^2 r_R \\
\ea \right \} .  \label{8}
\ee
\end{widetext}
Here, $ Z = 1 - \alpha_L^2 \alpha_R^2 r_L r_R. $

Solving the equations (\ref{2}), (\ref{3}) one arrives at the
following expression for the electron transmission \cite{36}:
\be
\tau (E) = \frac{J_2'}{J_1} = T_{21} + \sum_{m,n} K_m^{(2)}
(W^{-1})_{mn} K_n^{(1)}.    \label{9}
\ee
Within the considered model, $ 1 \leq m, \ n \leq 2, $
\begin{align}
K_m^{(1)} = T_{2m+1,1} + T_{2m +2,1},
\nn\\
K_m^{(2)} = T_{2,2m+1} + T_{2,2m +2},  \label{10}
\end{align}
and matrix elements of $ 2 \times 2 $ matrix $ W $ are given by
\be
W_{mn} = (2 - R_{mm})\delta_{mn} - \tilde R_{mn}(1 - \delta_{mn})  \label{11}
\ee
where
\begin{align}
R_{mm} =& T_{2m+1,2m+1} + T_{2m +2,2m+2}
\nn\\ &
+ T_{2m+2,2m+1} + T_{2m +1,2m+2},
 \nn\\
\tilde R_{mn} = &T_{2m+1,2n+1} + T_{2m +1,2n+2}
\nn\\ &
+ T_{2m+2,2n+1} + T_{2m +2,2n+2}. \label{12}
\end{align}
  It is known that thermoelectric efficiency of a thermoelectric
materials becomes reduced if the phonons thermal conductance takes on
significant values. In the expression of thermoelectric figure of
merit which characterizes the efficiency of energy conversion at small
temperature gradients,the denominator is the sum of electron and
phonon thermal conductances \cite{37,38}.
We remark that within the adopted model the phonon thermal conductance
through the junction is zero. This seems a reasonable assumption for
experiments give low values of phonon thermal conductance in several
molecular junctions \cite{39,40}. This may be attributed to the fact
that in many molecules the majority of vibrational transitions lie
above the range determined by thermal energy provided that relevant
temperatures take on values below room temperature \cite{15,41}.

        In the following analysis we focus on a significantly coupled system.
Provided that the dephasing reservoirs are detached from the bridge
$(\epsilon_L = \epsilon_R = 0)$ and the barriers separating the
electrodes from the bridge are identical $(t_L = t_R = t,\ r_L = r_R =
r), $ the electron transmission determined by Eqs.
(\ref{8})-(\ref{12}) accepts a simple form:
\be
\tau(E) = \frac{t^4}{(1 + r^2)^2} .  \label{13}
\ee   
 As known, in the case of coherent transport, the current flowing
through the system could be presented in the form:
 \begin{align}
I = &\frac{ie}{h} \sum_\sigma \int dE \big \{ \big(\Gamma_L^\sigma f_L
- \Gamma_R^\sigma f_R \big)\big(G_\sigma^r - G_\sigma^a \big)
 \nn \\ & +
\big(\Gamma_L^\sigma - \Gamma_R^\sigma \big) G_\sigma^<
    \big \}        \label{14}
\end{align}
 Here, $ \Gamma_{L,R}^\sigma (E)$ are self-energy terms  describing
coupling of an electron on the bridge (with a certain spin orientation
$\sigma $) to the electrodes, $ f_{L,R} $ are Fermi distribution
functions for the electrons on the electrodes, and $ G_\sigma^{r,a,<}
(E) $ are the retarded, advanced and lesser Green's functions
associated with the QD/molecule.

 Further, we accept a wide-band approximation for the self-energy
terms, and we concentrate on a symmetrically coupled system assuming $
\Gamma_L^\sigma = \Gamma_R^\sigma = \Gamma. $ Then the term
proportional to the lesser Green's function disappears from Eq.
(\ref{14}), and we obtain:
\be
 I = \frac{e}{\pi\hbar} \int \tau(E) (f_L - f_R)dE        \label{15}
\ee
where the electron transmission function is given by
\be
 \tau(E) = \frac{1}{2}\Gamma \sum_\sigma \big[G_\sigma^r (E) -
G_\sigma^a(e) \big] \equiv g^2(E).       \label{16}
\ee
 So, we have derived two expressions for the electron transmission
function appropriate to describe coherent electron transport through
the considered system. One of them (Eq. (\ref{13})) is obtained using
the scattering theory whereas another one  (Eq. (\ref{16})) is ensued
employing NEGF formalism. Within the considered coherent limit these
expressions should be identical. Comparing them, we get:
\be
t^2 = \frac{2g}{1 +g}. \label{17}
\ee
Provided that the electron transport through the junction is
undisturbed by electron-phonon interactions and disregarding spin-flip
processes, $ G_\sigma^r(E) $ may be approximated as \cite{42}:
\be
G_\sigma^r(E) = \frac{E - E_0 - \Sigma_{02}^\sigma - U\big(1-
\big<n_{-\sigma}\big>\big)}{(E - E_0 - \Sigma_{0\sigma})(E - E_0 - U -
\Sigma_{02}^\sigma) + U\Sigma_{1\sigma}} . \label{18}
\ee
In this expression, $ U $ is the charging energy describing Coulomb
repulsion between electrons on the bridge, and $ \big<n_\sigma\big>$
are one-particle occupation numbers which could be computed by
integration of the lesser Green's function $ G_\sigma^< (E) $ over the
whole range of tunnel energy $ E $ values:
\be
\big< n_\sigma  \big >  = -\frac{i}{2\pi}  \int G_\sigma^< (E)  dE.
             \label{19}
\ee
 Self-energy terms $ \Sigma_{o\sigma},\ \Sigma_{1\sigma} $ and $
\Sigma_{2\sigma} $ appear in the Eq. (\ref{10}) due to the coupling of
the bridge to the leads:
 \begin{align}
\Sigma_{0\sigma} = & \sum_{r\beta}\frac{|t_{r\beta;\sigma}|^2}{E -
\epsilon_{r\beta\sigma} + i\eta},    \label{20}
\\
\Sigma_{1\sigma} = & \sum_{r\beta} {|t_{r\beta;\sigma}|^2}
f_{r,-\sigma}^\beta \Big\{\frac{1}{E - \epsilon_{r\beta;-\sigma} +
i\eta}
\nn\\  & +
\frac{1}{E - 2E_0 - U +\epsilon_{r\beta;-\sigma} + i\eta} \Big\},    \label{21}
\\
\Sigma_{2\sigma} = & \sum_{r\beta} |t_{r\beta;\sigma}|^2
\Big\{\frac{1}{E - \epsilon_{r\beta;-\sigma} + i\eta}
\nn\\  & +
\frac{1}{E - 2E_0 - U +\epsilon_{r\beta;-\sigma} + i\eta} \Big\},
 \label{22}
\\
\Sigma_{02} = & \Sigma_{0\sigma} + \Sigma_{2\sigma}.   \label{23}
 \end{align}
Here, $ t_{r\beta;\sigma}$ are parameters describing the coupling of $
r,\beta $ electron states on the electrode $ \beta(\beta = L,R) $ to
the bridge state, $ \epsilon_{r\beta;\sigma} $ are single-electron
energies in the electrode $ \beta, \ f_{r\sigma}^\beta $ is the Fermi
distribution function for the energy  $ \epsilon_{r\beta;\sigma} ,$
chemical potential $ \mu_\beta $ and  temperature $ T_\beta $ and
$\eta $ is a positive infinitisemal parameter.
     Previously introduced coupling parameters $ \Gamma_{L,R}^\sigma $
are closely related to $ \Sigma_{0\sigma},$ namely: $ \Sigma_{0\sigma}
= \Sigma_{0\sigma}^L+ \Sigma_{0\sigma}^R,\ \Gamma_{L,R}^{\sigma} = - 2 \mbox{Im} \Sigma_{o\sigma}^{L,R}.$ We remark that $ \Sigma_{1\sigma} $
and $ \Sigma_{2\sigma} $ depend on the temperatures of electrodes
which is taken into account in further computations.

 Scattering probabilities $ \epsilon_{L,R} $ may be given an explicit
physical meaning by expressing them in terms of relevant energies. In
the considered system, dephasing and energy dissipation originate from
 interactions of charge carriers in the electrodes with thermal
phonons.  So, one can approximate these parameters as follows:
\be
\epsilon_\beta = \frac{\Gamma_{ph}^\beta}{\Gamma_\beta +
\Gamma_{ph}^\beta} . \label{24}
\ee
 Here, $ \Gamma_{ph}^\beta $ represents the self-energy term occurring
due to electron-phonon interactions in the electrode $\beta.$  Using
NEGF to compute the relevant electron and phonon Green's functions
within the self-consistent Born approximation, one may derive a
relatively simple approximation for $ \Gamma_{ph}^\beta$ \cite{43}:
\begin{align}
 \Gamma_{ph}^\beta = & \frac{2\pi}{\Gamma_\beta^2} \int_0^\infty
d(\hbar\omega) \sum_\alpha |\lambda_{\alpha\beta}|^2
\delta(\hbar\omega - \hbar\omega_{\alpha\beta}) \sum_{r\sigma}
|t_{r\beta;\sigma}|^2
          \nn\\ & \times
 \Big\{ \big[1 - f_\sigma^\beta(E - \hbar\omega)\big]\big[1 +
N_\beta(\omega)\big] \delta(E - \epsilon_{r\beta;\sigma} - \hbar\omega)
           \nn\\ & +
 f_\sigma^\beta(E - \hbar\omega) N_\beta(\omega) \delta(E -
\epsilon_{r\beta;\sigma} - \hbar\omega)
 \nn\\ & +
  f_\sigma^\beta(E + \hbar\omega) \big[1 + N_\beta(\omega)\big]
\delta(E - \epsilon_{r\beta;\sigma} + \hbar\omega)
 \nn\\ & +
 \big[1 - f_\sigma^\beta(E + \hbar\omega)\big] N_\beta(\omega)
\delta(E - \epsilon_{r\beta;\sigma} + \hbar\omega) \Big\} \label{25} .
 \end{align}
In writing down this expression, we had assumed that the coupling
strength $ \lambda_{\alpha\beta} $ characterizing interaction between
an electron in the electrode $ \beta $ and a thermal phonon with the
energy $ \hbar\omega_\alpha $ does not depend of the electron state $
r,\sigma. $
The electrodes are kept at different temperatures $T_\beta ,$  so we
introduce two phonon distribution functions $ N_\beta(\omega) = \big
\{\exp \big[\hbar\omega/kT_\beta\big] - 1 \big\}^{-1},\ ( k $ being
the Bolzmann's constant).

The phonon spectral function may be determined using molecular dynamic
simulations. However, to qualitatively analyze the effect of thermal
phonons on the transport characteristics, one may use the
approximation \cite{44}:
\begin{align}
\rho_{ph}^\beta(\omega) = & \sum_\alpha |\lambda_{\alpha\beta}|^2  \delta(\hbar\omega - \hbar\omega_{\alpha\beta})
        \nn\\ = &
\lambda_\beta \left(\frac{\omega}{\omega_{c\beta}}\right) 
e^{-\omega/\omega_{c\beta}} \theta(\omega)
  \label{26}
\end{align}
where $ \theta(\omega) $ is the Heaviside step function,  and $ \omega_{c\beta} $ characterize relaxation times for the thermal phonons.

 Substituting the approximation (\ref{26}) into Eq. (\ref{25}) one may
see that the major contribution to the integral comes from the region
where $ \omega \sim \omega_c. $ On these grounds,  one may replace $
\omega $ by $ \omega_{c\beta} $ in the arguments of slowly varying functions
in the integrand of Eq (\ref{25}).  Also, one may assume that $
\hbar\omega_{c\beta} $ is much smaller than $ \mu_\beta. $ So,
corrections including $ \hbar\omega_{c\beta} $ in the arguments of the
Fermi distribution functions and delta functions may be omitted. Then
one may use the expression for $ \Gamma_\beta^\sigma $ which follows
from its definition:
 \be
\Gamma_\beta^\sigma = 2\pi \sum_r |t_{r\beta;\sigma}|^2 \delta (E -
\epsilon_{r\beta;\sigma})   \label{27}
 \ee
to reduce Eq. (\ref{25}) to the form:
\be
 \Gamma_{ph}^\beta = \frac{2\lambda_\beta}{\Gamma_\beta}
\hbar\omega_{c\beta}
\left(\frac{kT_\beta}{\hbar\omega_{c\beta}}\right)^2  \zeta
\left(2;\frac{kT_\beta}{\hbar\omega_{c\beta}} + 1\right)
\label{28}
\ee
where $\zeta (x;q) $ is the Riemann's  $\zeta $ function. If  the
thermal energy $ kT_\beta $ significantly exceeds $ \hbar\omega_c, \
\Gamma_{ph}^\beta $ accepts an especially simple form, namely: $
\Gamma_{ph}^\beta \approx 2\lambda_\beta kT_\beta\big/\Gamma_\beta. $
In further calculations we assume for simplicity that $ \lambda_L =
\lambda_R = \lambda. $

For a considered symmetrically coupled junction, the thermocurrent is
described by the following expression:
\begin{align}
I_{th} = & \frac{e}{\pi\hbar} \int dE \big\{\tau(E,T_0,\Delta T)
f^L(E,T_0 + \Delta T)
\nn\\ & -
\tau(E,T_0,\Delta T =0) f^L(E,T_0) - \Delta\tau f^R(E,T_0) \big\}    \label{29}
\end{align}
where  $ \Delta \tau $ is given by:
\be
\Delta\tau = \tau(E,T_0,\Delta T) - \tau(E,T_0, \Delta T = 0).    \label{30}
\ee

The suggested approach allows one to analyze the effect of thermal
phonons on characteristics of thermoelectric transport beyond the
linear regime. Using Eqs. (\ref{8})-(\ref{28}), one may compute the
electron transmission function for an arbitrary value  of the ratio $
\Delta T/T_0 .$  This result may be used to calculate thermocurrent
and analyze how it is affected by various characteristics of the
considered junction (such as the quality of contact between the
electrodes and the linker, electron-electron and electron-phonon
interactions) and by external factors including the bias and gate
voltage and temperature gradient.

\section{iii. discussion}

The most interesting thermoelectric properties are better pronounced
in weakly coupled junctions where electron transmission exhibits sharp
maxima \cite{20}. Correspondingly, in further analysis we assume that
$ \Gamma < kT_0. $
  Then, as shown in the Fig. 2, $ \Delta \tau $ displays sharp dips at
$ E = E_0 $ and $ E = E_0 + U, $ and the magnitudes of these features
increase as the difference in the temperatures of electrodes enhances.
Electron-phonon interactions significantly affect the transmission. As
the coupling between electrons and thermal phonons strengthens, $
\Delta \tau $ generally takes on greater values. However, too strong
electron-phonon interactions bring partial spreading of the resonance
features which is not favorable for nonlinear behavior of
thermocurrent to be revealed.

\begin{figure}[t] 
\begin{center}
\includegraphics[width=8.8cm,height=4.3cm]{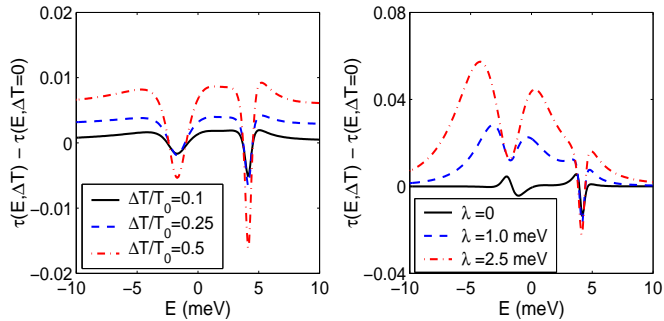}
\caption{(Color online) Energy dependencies of the electron
transmission through the junction. The function $ \Delta\tau =
\tau(E,\Delta T) - \tau(E,\Delta T = 0) $ is shown assuming $ E_0 =
-2meV,\ U = 6meV,\ \Gamma = 0.3meV,\ kT_0 = 0.65 meV $ at $\lambda =
0.5meV $(left panel) and $ \Delta T = 0.5 T_0 $  (right panel).
}
 \label{rateI}
\end{center}\end{figure}

Thermocurrent dependence of the applied bias voltage is illustrated in
the left panel of the Fig. 3 assuming for certainty that the left
(hot) electrode is kept at higher voltage and the bridge energy level
$ E_0 $ is situated below the Fermi level for the electrodes. When the
bias is small so that  $E_0 $ remains outside the conduction window
whose width is determined by the difference between chemical
potentials of electrodes  $ \mu_L $ and $\mu_R ,$ the charge flow is
driven by the temperature gradient. Higher temperature of the left
electrode enhances probability for an electron to tunnel there  from
the bridge provided that $ E_0 $ is rather close to the Fermi level,
so that their difference is of the same order as the thermal energy $
kT_0. $ Under these  conditions, $ I_{th} $ takes  on positive values.
When the bias becomes greater, $ E_0 $ enters the conduction window,
and the electric driving forces come into play. As a result, the
thermocurrent changes its sign. It remains negative at moderate bias.
In strongly biased junctions, electric driving forces predominate, and
$ I_{th} $ approaches zero. The depth of the dip appearing on the $
I_{th} - V $ curves, as well as its position, are controlled by
several factors. Assuming that the charging energy $ U $ and
electron-phonon coupling strength $ \lambda $ are fixed, it is
determined by the value of $ \Delta T. $ The greater becomes the
difference between the electrodes temperatures, the greater is the
maximum magnitude of $ I_{th} $ indicating more favorable conditions
for energy conversion in the considered system.

\begin{figure}[t] 
\begin{center}
\includegraphics[width=8.8cm,height=4.3cm]{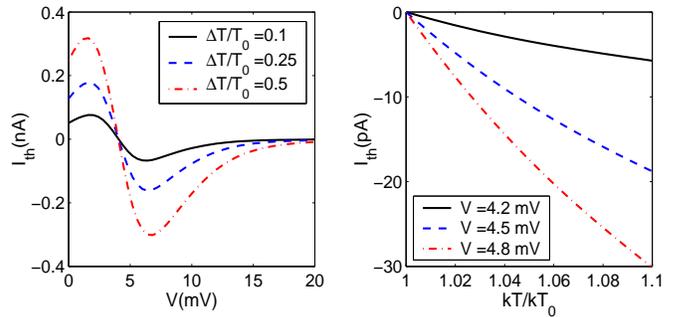}
\caption{(Color online)  Left panel: Thermocurrent as the function of
bias voltage. The curves are plotted at $ U = 6meV,\ \lambda =
0.5meV,\ kT_0 = 0.65meV. $  Right panel: Temperature dependencies of
the  thermocurrent. The curves are plotted assuming $ kT_0 = 0.65meV,\
 \lambda = 0.5meV. $ Remaining parameters are the same as in the Fig.
2.
}
 \label{rateI}
\end{center}\end{figure}

It was  reported \cite{19} that thermocurrent flowing through
a quantum dot may exhibit a nonlinear dependence of $ \Delta T $ even
at small values of the latter. As recently suggested \cite{25}, the
nonlinearity appears due to renormalization of the energy $ E_0 $ in
the presence of temperature gradient. Taking into account the
suggested energy renormalization, we showed that weak nonlinearity of
$ I_{th} $ versus $ \Delta T $ curves may be actually traced (see
right panel of the Fig. 3). Presented curves demonstrate  a
qualitative agreement with the experimental results of Ref. \cite{19}.
However, we remark that the suggested renormalization $ E_0 \to E_0 +
zk\Delta T $ where $|z| < 1 $ may significantly affect the
thermocurrent behavior only at certain values of the bias voltage,
when $ E_0 $ is very close to the boundary of the conduction window.

Coulomb repulsion between electrons on the bridge and the electron
interactions with thermal phonons  may significantly influence
thermoelectric transport through the junction, as demonstrated in the
Fig. 4. The enhancement of charging energy narrows down the interval
where $ I_{th} $ accepts negative values and makes the dips in the $
I_{th} - V $ curves more shallow. This may be explained by considering
the role taken by Coulomb interactions. Electron-electron interactions
are hindering electron flow through the system in both directions,
which  leads to the Coulomb blockade. These interactions could be
treated as a source of an effective force opposing any predominating
driving force (originating either from the bias voltage or from the
temperature gradient applied across the junction). As a result, the
characteristic features manifested in the shapes  of $ I_{th} - V $
curves become less distinctly pronounced.

Electrons interactions with thermal phonons do not change the width of
the region where $ I_{th} $ remains negative. However, these
interactions may affect the magnitude of thermocurrent. In the right
panel of the Fig. 4, we present the  $ I_{th} $  value at the bottom
of the dip as a  function of the electron-phonon coupling strength. As
shown in the figure, the dip depth reduces as $ \lambda $ increases.
This effect is better pronounced at weak or moderate electron-phonon
coupling $(\lambda \lesssim \Gamma, kT_0),$ and it fades away when  $
\lambda $ significantly exceeds  $ \Gamma. $

\begin{figure}[t] 
\begin{center}
\includegraphics[width=8.8cm,height=4.3cm]{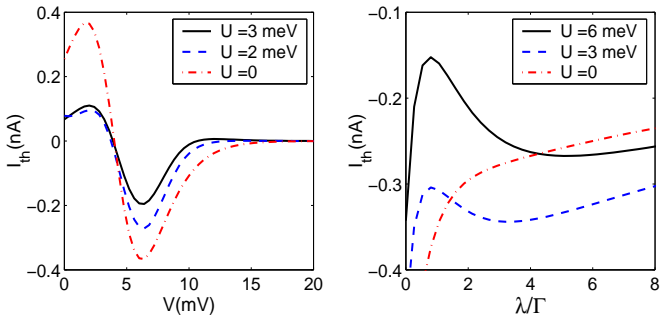}
\caption{(Color online) The effect of electron-electron (left panel)
and electron-phonon (right panel) interactions on the thermocurrent.
Curves are plotted at $ \Delta T = 0.25 T_0,\ \lambda = 0 $ (left
panel) and at $ V = 6meV $ (right panel).  Other relevant parameters
are the same as in the previous figures.
}
 \label{rateI}
\end{center}\end{figure}

\begin{figure}[t] 
\begin{center}
\includegraphics[width=8.8cm,height=4.3cm]{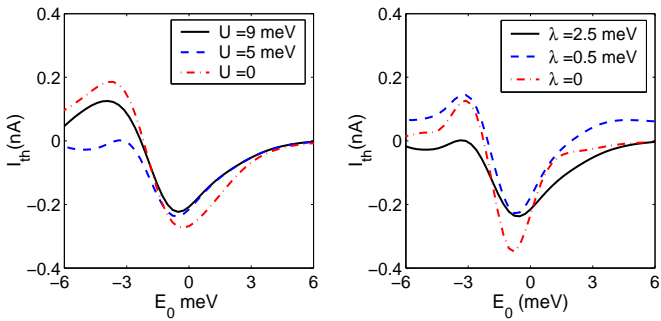}
\caption{(Color online) Thermocurrent as a function of the bridge
level position $ E_0. $ Curves are plotted assuming $ V = 4mV,\ \Delta
T = 0.25 T_0,\ \lambda = 2.5 meV $ (left panel) and  $ U = 5meV $
(right panel). Other relevant parameters are the same as in the
previous figures.
  }
 \label{rateI}
\end{center}\end{figure}
As known, transport properties of molecular junctions and other
similar  systems may be controlled by varying the positions of the
bridge energy levels. Practically, the levels may be shifted by a gate
voltage applied to the system. Within the accepted model, the bridge
in the considered junction is represented by a single energy level $
E_0. $ In the Fig. 5 we trace the thermocurrent  dependencies of  $
E_0. $ The presented results confirm those shown in the previous
figures. Again, the thermocurrent acquires a negative sign when the
bridge level moving upwards appears in the conduction window. For a
symmetrically coupled system this happens at $E_0 = - \frac{1}{2}V $
assuming that the Fermi level for unbiased junction $ \mu = 0. $As the
bridge level moves higher, $ I_{th} $ remains negative while the level
is still inside the conduction window. When it leaves the  window but
remains near its upper boundary, the thermocurrent may change sign
once more, being influenced by electron-electron interactions and
electron-phonon interactions in the electrodes. However, when the
bridge level moves farther away from the conduction window, both terms
in the Eq. (\ref{1}) approach zero, so the thermocurrent disappears.

\section{iv. conclusion}

Finally, we repeat again that various aspects of energy conversion in
nanoscale systems attract significant interest of the research
community. The present work was inspired by this common interest.
Also, the present research was motivated by recent experimental
observations of nonlinear thermocurrent flowing through quantum dots.
The thermocurrent defined by Eq. (\ref{1}) is an important
characteristic of thermoelectric  transport. The behavior of the thermocurrent brings some information concerning the efficiently of energy conversion in the considered  systems. For instance, $ I_{th} $ reaches minimum at a certain value of the applied bias voltage, as shown in the Figs. 3,4. One may conjecture that this minimum occurs when the current $
I(V,T_0, \Delta T) $ becomes zero, so the corresponding magnitude of
bias voltage is close to the magnitude of thermovoltage $ V_{th} $ and
may be employed to estimate the latter. However, we remark that one
cannot extract sufficient information to properly estimate the
thermoelectric efficiency of the system basing on the thermocurrent
behavior alone. Even within the linear  in temperature regime one
needs additional information concerning electron electrical and
thermal conductances and  thermopower to make such estimates.

In a nanoscale junction consisting of a quantum dot or molecule
sandwiched in between two electrodes, the thermocurrent value is
controlled by several factors. These include temperatures of
electrodes, bias and gate voltage, charging energy characterizing
electron-electron interactions on the bridge and specific energies
characterizing the coupling of the bridge to the electrodes and
electron-phonon interactions. To theoretically analyze possible effect
of the above factors on thermocurrent we employ a single-particle
scattering approach pioneered by Landauer in the context of charge
transport in mesoscopic systems.

For simplicity, we simulate the bridge  in the considered junction by
a single orbital, and we assume that it  is symmetrically coupled  to
the electrodes. Then the thermocurrent  is expressed in terms of
electron transmission functions (which depend on all above mentioned
factors) and Fermi distribution functions for the electrodes. We
derive the expression for the electron transmission which  remains
valid for an arbitrary value of the difference between the
temperatures of electrodes. This makes it suitable for analysis of
thermoelectric transport both within and beyond linear in temperature
regime. Using this expression, we show that varying the bias and gate
voltage one may create favorable conditions for heat-to-electricity
conversion in a junction assuming that the most important
characteristic energies $ kT_0,\ U $ and $ \lambda $ are fixed. Also,
we analyze how electron-electron interactions on the bridge and
electrons interactions with thermal phonons associated with electrodes
may affect the thermoelectric properties of the considered systems.
Coulomb repulsion between electrons opposes electron transport through
the junction at small bias voltage. We show that this brings a partial
suppression of the thermocurrent. Electron-phonon interactions in the
electrodes assist in the increase of scattering probabilities thus
destroying the coherence of electron transport and bringing additional
suppression of thermocurrent, as illustrated in Figs. 4,5.

The computational method employed in the present work may be further
generalized to include the effect of molecular vibrations. Also, one
may simulate the bridging molecule/quantum dot by several orbitals
thus opening  the way to studies of quantum interference effects. So,
we believe that presented computational scheme and obtained results
may  be helpful for further understanding of thermoelectric properties
of nanoscale systems.

\section{ Acknowledgments}
 {\bf Acknowledgments:}
 This work was  supported  by  NSF-DMR-PREM 0934195. The author
thanks  G. M. Zimbovsky for help with the manuscript.

\end{document}